\pdfoutput=1
\documentclass[journal]{IEEEtran}
\ifCLASSINFOpdf
\else
\fi
\usepackage{amsmath}
\usepackage{amsmath}
\usepackage{amssymb} 
\usepackage{graphicx}
\usepackage{caption}
\usepackage{subcaption}
\usepackage{bm}
\usepackage{mathrsfs}
\usepackage{pdfrender}
\usepackage{tikz}
\usepackage{graphicx} % For including figures
\usepackage{changepage} % For adjusting margins locally
\usepackage{algorithm}
\usepackage{algpseudocode}
\usepackage{algorithm}
\usepackage{algpseudocode}
\usepackage{booktabs} % For professional-looking tables
\usepackage{array}    % For m{} column type
\usepackage{caption}  % For customizing captions
\usepackage{cite}
\usepackage{bbm}
\usepackage{mathtools} % for \mathclap
\usepackage{algorithm}
\usepackage{algpseudocode}
\usepackage{amsmath}
\usepackage{amssymb}
\usepackage{algorithmicx}
\usepackage{tikz}
\usetikzlibrary{positioning,arrows.meta}
\usetikzlibrary{shapes.geometric}

\newcommand{\mathoutline}[1]{%
  \text{\pdfrender{StrokeColor=black, TextRenderingMode=1, LineWidth=0.08pt}{\ensuremath{#1}}}%
}

\begin{document}

\title{Sparse Operator-Adapted Wavelet Decomposition Using Polygonal Elements for Multiscale FEM Problems\\}

%% \title{Near-Linear Multiscale FEM on Unstructured Arbitrary Polygonal Meshes\\}

%% \title{Sparse Operator-Adapted Wavelets for Multiscale FEM on Unstructured Polygonal Mesh Hierarchies\\}


\author{F.~\c{S}\i k,~\IEEEmembership{Graduate Student Member,~IEEE,}
        F.~L.~Teixeira,~\IEEEmembership{Fellow,~IEEE,}
        and~B.~Shanker,~\IEEEmembership{Fellow,~IEEE}% <-this % stops a space
\thanks{The authors are with the Department of Electrical and Computer Engineering and Electroscience Laboratory, The Ohio State University, Columbus, OH 43210, USA}% <-this % stops a space
\thanks{Manuscript received MM DD, YYYY; revised MM DD, YYYY.}}

\maketitle


\begin{abstract}
We develop a sparse multiscale operator-adapted wavelet decomposition-based finite element method (FEM) on unstructured polygonal mesh hierarchies obtained via a coarsening procedure. Our approach decouples different resolution levels, allowing each scale to be solved independently and added to the entire solution without the need to recompute coarser levels. At the finest level, the meshes consist of triangular elements which are geometrically coarsened at each step to form convex polygonal elements. Smooth field regions of the domain are solved with fewer, larger, polygonal elements, whereas high-gradient regions are represented by smaller elements, thereby improving memory efficiency through adaptivity. The proposed algorithm computes solutions via sequences of hierarchical sparse linear-algebra operations with nearly \(\mathcal{O}(N)\) computational complexity. 
\end{abstract}


% \fi
\IEEEpeerreviewmaketitle

\section{Introduction}
The finite element method (FEM) is an accurate and flexible approach to solving electromagnetic problems in complex geometries~\cite{jin2015tcef,jin2015finite,peterson1998computational}.  Adaptive FEM approaches are often used to locally resolve fine details of the solution by refining the mesh only over selected regions of the domain~\cite{szabo1991fea,sun2000amr}. However, the resulting FEM matrices couple different resolution levels, which incurs undesired computational overhead. Scale coupling also worsens the condition number of the FEM matrices, thereby slowing the iterative solvers. Scale-coupling remains a significant limitation of adaptive FEM approaches for multiscale problems~\cite{filippi2015performance,filippi2020electromagnetic,xu2020ddmfe,sudarshan2005operator}. 

To overcome this challenge, a range of multiscale techniques have been explored. Some early studies employed first-generation (bi-orthogonal) wavelets as basis functions, capitalizing on their natural multiscale representation. Although first-generation wavelets improved sparsity in some settings, finite element applications still retained scale coupling, and thus did not deliver significant gains~\cite{mallat1989theory,daubechies1990wavelet,williams1994introduction,wagner1995wavelets}. Subsequent efforts turned to second-generation wavelets that yield scale-decoupled systems for certain classes of problems; however, generalization across mesh types, geometries, and boundary conditions has proved challenging, and increases in overall computational costs have been reported~\cite{amaratunga2006multiresolution,sudarshan2006combined,quraishi2011second,lounsbery1997mra}. 

More recently, operator-adapted wavelet decomposition-based finite element approaches have been developed to produce scale-decoupled systems~\cite{sudarshan2005operator,budninskiy2019operator,sik2025multiscale}. Using this approach, full decoupling among the coarse and all detail levels is obtained. Consequently, deep-level detail can be added to the total solution without recomputing coarser level ones. Such systems were first demonstrated for simple geometries discretized on {\it structured} meshes~\cite{budninskiy2019operator}. To make this approach applicable to more general problems, we have recently extended the sparse operator-adapted wavelet decomposition-based FEM to {\it irregular unstructured meshes}
in~\cite{sik2025oaw,sik2025multiscale} while retaining
near-\(\mathcal{O}(N)\) computational complexity. 
This paper extends and generalizes that work by incorporating a new coarsening mesh hierarchy based on {\it polygonal} meshes while maintaining accuracy and near-linear time complexity.
The coarsening proceeds from an adaptively generated finest-level triangulation to progressively larger \emph{convex polygonal} elements (e.g., quadrilaterals, pentagons, hexagons) as illustrated in Fig.~1. In practice, smooth regions are covered by fewer elements (polygons), while regions with high field gradients and singularities retain smaller elements. This strategy provides an effective unification of a multiscale FEM formulation with localized $h$-adaptivity to produce small scale-decoupled FEM matrices at each detail level. 
The main contributions of this paper can be summarized as follows.
\begin{itemize}
  \item Extension of operator-adapted wavelets to unstructured polygonal mesh hierarchies via a mesh-coarsening algorithm that generates arbitrary convex polygonal elements at each scale level, thereby combining the benefits of FEM multiscale modeling with $h$-adaptivity.
  \item Efficient construction of precomputed sparse operator-agnostic FEM matrices on convex polygonal meshes using edge elements based on generalized barycentric coordinates.
  \item Construction of memory-efficiency sparse FEM linear solvers  on unstructured convex polygonal mesh hierarchies with near-linear complexity.
\end{itemize}

The remainder of the paper is organized as follows. Section~\ref{sec:theory} presents the problem statement, a brief overview of multiresolution analysis, and the key properties of operator-adapted wavelets. Section~\ref{sec:opFEM} describes the hierarchical mesh coarsening procedure, the construction of precomputed operator-agnostic input matrices, the implementation of edge elements on convex polygonal elements, and the algorithm implementation outline. Section~\ref{sec:res} describes multiscale numerical experiments used to demonstrate the main features of the proposed method. Finally, Section~\ref{sec:conclusion} closes out the paper with some key conclusions.

\begin{figure*}[!t]
  \centering
  \includegraphics[width=\textwidth]{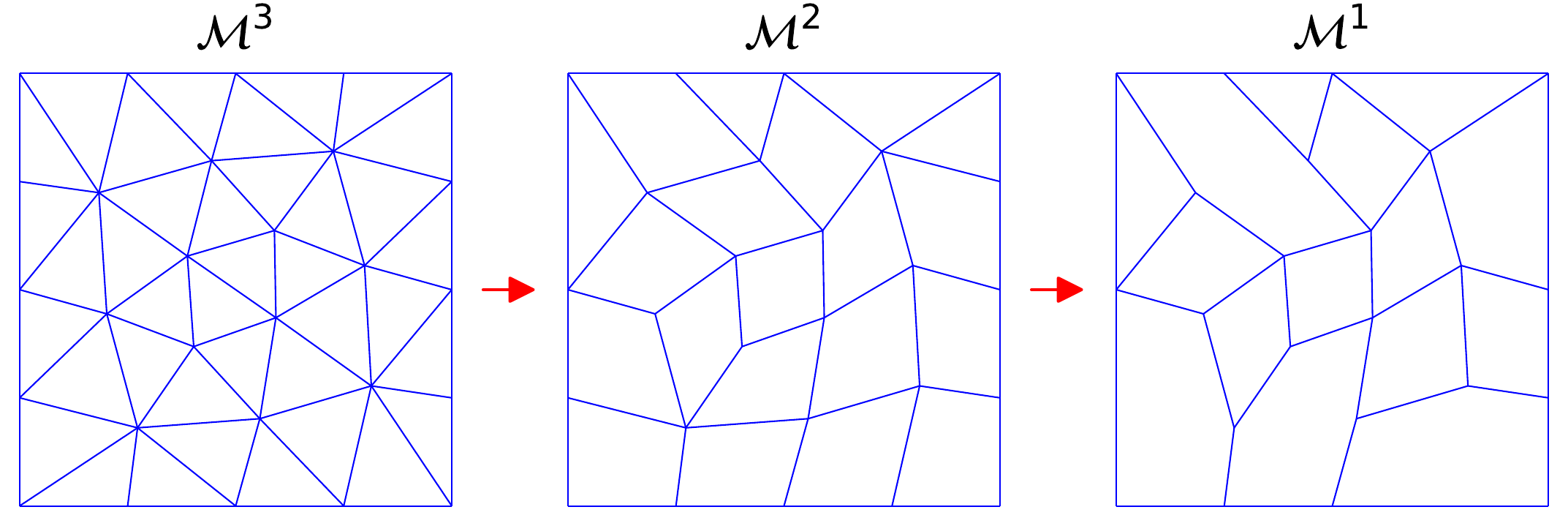}
  \caption{Illustration of a three-level mesh hierarchy produced by the proposed coarsening algorithm. From left to right: $\mathcal{M}^{3}$ (input finest-level uniform mesh) with 38 triangular elements; 
  $\mathcal{M}^{2}$ (obtained by coarsening the finest mesh once) with 18 quadrilaterals and 2 triangles; and $\mathcal{M}^{1}$ (coarsest-level mesh) with 4 hexagons, 1 pentagon, 9 quadrilaterals, and 1 triangle.}
  \label{fig:mesh-hierarchy}
\end{figure*}

\section{Background\label{sec:theory}}
\subsection{Problem Statement and FE Discretization}

 We consider the vector Helmholtz equation for \(\mathbf{u}(\mathbf{r})\), representing either the electric or magnetic field intensity:
\begin{equation} \label{eq:eq1}
{\mathcal{L}} \,\mathbf{u}(\mathbf{r})=
\left[ \nabla \times \left( \frac{1}{\alpha_1}\,\nabla \times \right) -\alpha_2 \, k_0^2  \right]\,\mathbf{u}(\mathbf{r}) = \mathbf{g}(\mathbf{r}) \quad \text{in } \Omega,
\end{equation}
with boundary conditions
$\mathcal{B}_i\{\mathbf{u}(\mathbf{r})\} = \mathbf{b}_i(\mathbf{r})$ on $\Gamma_i$, where \(\mathcal{B}_i\) is a generic differential operator and \(\mathbf{b}_i\) the boundary data. For electric field equation \((\alpha_1,\alpha_2)=(\mu_r,\epsilon_r)\), whereas for magnetic field \((\alpha_1,\alpha_2)=(\epsilon_r,\mu_r)\). Here, \(\mathbf{g}(\mathbf{r})\) denotes the source term and \(k_0\) is the free-space wavenumber.

The proposed hierarchical coarsening–based algorithm is applied to a two-dimensional (2D) domain \(\Omega\) discretized by arbitrary convex \textit{polygonal} elements (triangles, quadrilaterals, pentagons, hexagons, etc.). Each element \( \mathbf{u}^e(\mathbf{r}) \) occupies a subdomain \(\Omega^e\) for \(e=1,2,\ldots,n^e\), where \(n^e\) denotes the total number of elements, and expanded as a weighted sum of Whitney one-form basis functions (edge elements) \(\mathbf{W}_j^e(\mathbf{r})\) defined on convex polygonal elements such that \(\mathbf{W}_j^e(\mathbf{r}) \in H(\mathrm{curl}, \Omega^e)\):
\begin{equation}
\mathbf{u}^e(\mathbf{r}) = \sum_{j=1}^{s^e} u_j^e\, \mathbf{W}_j^e(\mathbf{r}),
\end{equation}
where \(u_j^e\) are the elemental degrees of freedom associated with the \(j\)th edge, \(s^e\) is the number of local DoFs (equal to the number of edges of the element, \(s^e=3\) for triangles, \(4\) for quadrilaterals, \(5\) for pentagons, etc.), Details on the assignment of Whitney basis functions to convex polygonal elements with an arbitrary number of edges are provided in Section~\ref{sec:opFEM}.

\subsection{Operator-Adapted Multiresolution Analysis}

Wavelet multiresolution analysis offers a hierarchical framework for approximating solutions to multiscale electromagnetic problems.
In traditional wavelet–Galerkin formulations, finite-energy orthogonal wavelet bases with compact support are used to discretize (1). The solution is approximated in a finite-dimensional space \(V^q\) built within a nested sequence of approximation spaces \(\{V^j\}_{j=1}^{q-1}\subset H\):
\begin{equation}
\{0\}\subset V^1\subset\cdots\subset V^{j-1}\subset V^{j}\subset V^{j+1}\subset\cdots\subset V^q.
\end{equation}
For each level \(j\), there exists a complementary wavelet (detail) space \(W^j\) that supplies the information needed to pass from \(V^j\) to \(V^{j+1}\), and
\begin{equation}
V^{j+1}=V^j\oplus W^j,\qquad j=1,2,\dots,q-1,
\end{equation}
where \(\oplus\) denotes the direct sum in the \(L^2\) space.
However, in a traditional multiresolution approach, as evident from the global FEM matrix \( \mathbf{L} \) in (5), different resolution levels are inherently coupled to each other. Due to this undesired scale coupling, adding finer detail levels to improve accuracy affects the coarser level solutions. This means that previously computed coarser levels must be recomputed together with each new detail level. 

The sparse operator-adapted wavelet decomposition–based FEM solves this issue by constructing a fully scale-decoupled system. As illustrated in (5), after applying the proposed approach, all inter-level couplings are eliminated and the global FEM matrix \(\mathbf{L}\) becomes block diagonal by scale.
Consequently, once the coarse-level solution has been computed, any number of finer detail levels can be added to the total solution as described in (6) via independent linear solves; the process can be terminated as soon as the prescribed accuracy is reached.

\begin{equation}
\begingroup
\small
\setlength{\arraycolsep}{3pt}
\renewcommand{\arraystretch}{1.1}
\mathbf{L}=
\underbrace{\left[
\begin{array}{ccc}
A^{1} & M_{\phi\psi}^{(1,1)} & M_{\phi\psi}^{(1,2)}\\
M_{\psi\phi}^{(1,1)} & B^{1} & M_{\psi\psi}^{(1,2)}\\
M_{\psi\phi}^{(2,1)} & M_{\psi\psi}^{(2,1)} & B^{2}
\end{array}
\right]}_{\mathclap{\scriptsize
\begin{array}{c}
\text{Scale-coupled system}\\[-2pt]
\text{(traditional wavelet--Galerkin)}
\end{array}}}
\;\Rightarrow\;
\mathbf{L}=
\underbrace{\left[
\begin{array}{ccc}
A^{1} & 0 & 0\\
0 & B^{1} & 0\\
0 & 0 & B^{2}
\end{array}
\right]}_{\mathclap{\scriptsize
\begin{array}{c}
\text{Scale-decoupled system}\\[-1pt]
\text{(operator-adapted wavelets)}
\end{array}}}
\endgroup
\label{eq:scale-coupling}
\end{equation}

\begin{equation}
\mathbbm{u}^{q}
= \mathbbm{u}^{\text{coarse}}
  + \sum_{j=1}^{q-1} \mathbbm{u}^{\text{detail},\,j}
\label{eq:u-composition}
\end{equation}

This scale decoupling is a consequence of the operator-orthogonality (\(\mathcal{L}\)-orthogonality) of operator-adapted wavelets \cite{budninskiy2019operator, owhadi2019operator, chen2019material, sik2025oaw, sik2025multiscale}. The operator-adapted wavelet space \(\mathbb{W}^j\) at any level \(j\) is \(\mathcal{L}\)-orthogonal to the solution space at its own level and to the wavelet spaces at all levels. The traditional multiresolution analysis (MRA) can be reformulated in the operator-orthogonality-based framework as:

\begin{equation}
\mathbb{V}^{j+1}=\mathbb{V}^{j}\oplus_{\mathcal{L}}\mathbb{W}^{j}, \quad j=1,2,\dots,q-1,
\end{equation}
for q-level system, where \(\oplus_{\mathcal{L}}\) denotes the direct sum of \(\mathcal{L}\)-orthogonal subspaces. Using this operator-orthogonality-based hierarchy, the finest-level space \(\mathbb{V}^{q}\) can be expressed as:
\begin{equation}
H \approx \mathbb{V}^{q}
  = \mathbb{V}^{1}
    \oplus_{\mathcal{L}} \mathbb{W}^{1}
    \oplus_{\mathcal{L}} \cdots
    \oplus_{\mathcal{L}} \mathbb{W}^{q-1}.
\end{equation}

Based on (7) and (8), an operator-orthogonality-based multiresolution scheme can be constructed to obtain the proposed scale-decoupled multiscale solver.

\subsection{Key Properties of Operator-Adapted Wavelets}

In this subsection, the key properties of operator-adapted wavelets are summarized. Comprehensive details are provided in our earlier paper~\cite{sik2025oaw}.

\begin{itemize}
  \item \textit{Notational conventions:} In this paper, we use a notational distinction between precomputed {\it operator-agnostic} and {\it operator-adapted} matrices/vectors. Precomputed operator-agnostic inputs (those obtained from the usual FEM assembly) are denoted by bold letters with a tilde (e.g., \(\tilde{\mathbf{A}}^{j}\)). In contrast, operator-adapted matrices and vectors are indicated by outlined fonts (e.g., \(\mathbb{A}^{j}\)).

  \item \textit{Operator-orthogonality:} As noted earlier, the operator-adapted wavelet space at resolution level \(j\), denoted \(\mathbb{W}^{j}\), is \(\mathcal{L}\)-orthogonal to the solution space \(\mathbb{V}^{j}\) at the same scale level. Here, \(\mathbb{W}^{j}\) is spanned by the operator-adapted wavelets \( \{ \mathoutline{\psi}_\ell^j \}_{\ell=1}^{N_j}\), while \(\mathbb{V}^{j}\) is spanned by the operator-adapted basis functions \( \{ \mathoutline{\phi}_i^j \}_{i=1}^{n_j}\). These spaces satisfy
  \begin{equation}
  \int_{\Omega}\mathoutline{\phi}_{i}^{\,j}\,\mathcal{L}\,\mathoutline{\psi}_{\ell}^{\,j,T}\,d\Omega \;=\; 0,
  \qquad \forall\, i,\ell,
  \label{eq:L-orthogonality}
  \end{equation}
  and, in addition, each \(\mathbb{W}^{j}\) is \(\mathcal{L}\)-orthogonal not only to \(\mathbb{V}^{j}\) but also to all other wavelet spaces \(\mathbb{W}^{m}\) with \(m \neq j\).

  \item \textit{Canonical multiresolution analysis:} A central requirement is that the basis functions (both operator-agnostic and operator-adapted) form a refinable hierarchy. Specifically, any coarse-level basis function \(\tilde{\boldsymbol{\phi}}_{i}^{j}\) can be represented as a weighted sum of basis functions at the next finer level \(\tilde{\boldsymbol{\phi}}_{\ell}^{\,j+1}\); the operator-adapted basis functions produced by our algorithm satisfy an analogous property.
  \begin{equation}
  \begin{aligned}
    \tilde{\mathbf{\phi}}_{i}^{j} &= \sum_{\ell=1}^{n_{j+1}} \tilde{\mathbf{C}}_{i\ell}^{\,j}\,\tilde{\mathbf{\phi}}_{\ell}^{\,j+1}
      \quad \text{(operator-agnostic basis)},\\[1ex]
    \mathoutline{\phi}_{i}^{j} &= \sum_{\ell=1}^{n_{j+1}} \mathbb{C}_{i\ell}^{\,j}\,\mathoutline{\phi}_{\ell}^{\,j+1}
      \quad \text{(operator-adapted basis)}.
  \end{aligned}
  \end{equation}
  
  \item \textit{Sparse precomputed operator-agnostic matrices and initialization of the main algorithm:} The proposed sparse operator-adapted wavelet decomposition algorithm is initialized using the usual FEM quantities at the finest level \(q\): we set \(\mathbb{A}^{q}=\tilde{\mathbf{A}}^{q}\) (system matrices), \(\mathoutline{\Phi}^{q}=\tilde{\bm{\Phi}}^{q}\) (edge basis functions), and \(\mathbbm{g}^{q}=\tilde{\mathbf{g}}^{q}\) (right-hand side). In addition, the matrices \(\tilde{\mathbf{C}}^{j}\) and \(\tilde{\mathbf{W}}^{j}\), whose construction will be described later, are precomputed at each scale level. All of these inputs are generated during precomputation steps using Whitney one-form basis functions.
  
   In our previous implementation~\cite{sik2025oaw,sik2025multiscale}, a subdivision-based strategy was used to construct the mesh hierarchy. In this work, we instead construct the mesh hierarchy by coarsening a given finest-level triangular (tetrahedral in 3D) element mesh, which is more representative of real-world practical scenarios. The coarse meshes inherently contain convex polygonal elements. At each level, operator-agnostic matrices are assembled by assigning edge basis functions to these polygonal elements and evaluating the corresponding integrals. The mathematical details for generating these precomputed inputs are presented in Section~\ref{sec:opFEM} where we emphasize the role of the precomputation stage and the aspects that differ in this work.

\end{itemize}

\begin{figure}[t]
\centering
\resizebox{0.57\columnwidth}{!}{%
\begin{tikzpicture}[node distance=7mm]
  % Styles
  \tikzset{
    flow/.style={draw=red!80!black, very thick, rounded corners=2mm,
                 inner sep=2.5pt, align=center},
    flowwide/.style={flow, minimum width=4.6cm},
    term/.style={flow, ellipse, inner sep=1.2pt},
    dblarr/.style={double, line width=0.5pt, -{Stealth[length=2.4mm]}}
  }

  % Title (optional: comment out if you prefer only the caption)
  % \node[font=\bfseries] at (0,1.2) {Coarsening Procedure};

  % Nodes
  \node[flow]      (s1) {Detect shared edges};
  \node[flowwide, below=of s1] (s2) {Merge candidate polygons};
  \node[flow,      below=of s2] (s3) {Convexity check};
  \node[flow,      below=of s3] (s4) {Accept merged polygons\\if the new polygon is convex};
  \node[flow,      below=of s4] (s5) {Otherwise\\keep original polygons};
  \node[term,      below=7mm of s5] (end) {END};

  % Arrows
  \draw[dblarr] (s1) -- (s2);
  \draw[dblarr] (s2) -- (s3);
  \draw[dblarr] (s3) -- (s4);
  \draw[dblarr] (s4) -- (s5);
  \draw[dblarr] (s5) -- (end);

  % Loop back from “Accept…” to top
  \draw[dblarr] (s4.west) -- ++(-1.4,0) |- (s1.west);
\end{tikzpicture}%
}

\caption{Mesh coarsening procedure.}
\label{fig:coarsen-flow}
\end{figure}
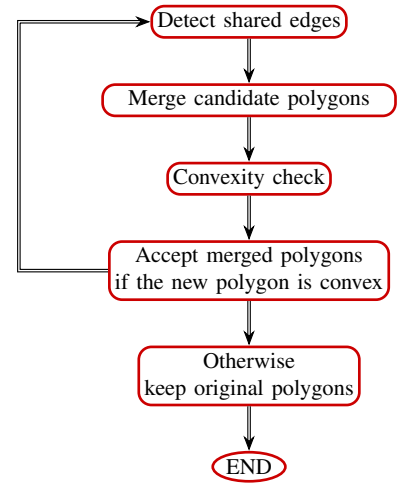

\section{Coarsening-Based Hierarchical Operator-Adapted Wavelet Decompositions\label{sec:opFEM}}

\subsection{Adaptive Multiscale Mesh Coarsening Procedure}

This subsection presents the mesh coarsening algorithm that generates the multiscale mesh hierarchy. Mesh generation starts from the finest-level ($q$-level) triangular tessellation \(\mathcal{M}^{q}\) . The finest mesh can be either uniform or adaptive as the proposed coarsening algorithm is compatible with both.  When \(\mathcal{M}^{q}\) is adaptive, with finer triangulation near high-field gradient regions and coarser triangulation in smooth regions, the benefits of the multiscale formulation combine with those of adaptivity: smooth field regions are covered by fewer larger elements; the total element count is reduced compared with a uniformly fine mesh. In a typical workflow, only the finest-level triangulation \(\mathcal{M}^q\) is assumed fixed a priori as given by an external mesh generation tool.
Coarser meshes \(\mathcal{M}^{q-1},\ldots,\mathcal{M}^{1}\) are automatically generated in succession by Algorithm~1, also illustrated in Fig.~2. During coarsening, candidate elements are merged to form larger cells while enforcing convexity. The resulting meshes thereby consist of a mix of convex polygons. 

\algrenewcommand\algorithmicrequire{\textbf{Inputs:}}
\algrenewcommand\algorithmicensure{\textbf{Outputs:}}

\begin{algorithm}
\caption{Hierarchical Mesh Coarsening Procedure}
\begin{algorithmic}

\Require Finest-level triangular mesh $\mathcal{M}^{q}=(\mathcal{V}^{q},\mathcal{E}^{q},\mathcal{P}^{q})$
\Ensure Mesh hierarchy $\{\mathcal{M}^{j}\}_{j=q}^{1}$ of \emph{convex} polygonal meshes

\vspace{0.35em}
\noindent\textbf{\textit{Notation.}}
At level $j$, $\mathcal{M}^{j}=(\mathcal{V}^{j},\mathcal{E}^{j},\mathcal{P}^{j})$ with
$\mathcal{V}^{j}$ vertices, $\mathcal{E}^{j}$ edges, and $\mathcal{P}^{j}$ cells (polygons).
For an interior edge $e=\{u,v\}\in\mathcal{E}^{j}$, let $m(e)=\{p_a,p_b\}\subset\mathcal{P}^{j}$ be its two adjacent cells.
A cell is \emph{free} if it has not been merged in the current pass.

\vspace{0.35em}
\For{$j = q$ \textbf{down to} $2$}
  \State Build the edge–to–cell map $m(e)$ for all $e\in\mathcal{E}^{j}$.
  \State Mark all $p\in\mathcal{P}^{j}$ as free; set $\texttt{merged}\gets 0$.
  \For{each interior edge $e=\{u,v\}$ with $m(e)=\{p_a,p_b\}$ and both $p_a,p_b$ free}
     \State \textit{Form candidate boundary} $B$:
         \Statex \quad Remove edge $e$ from the union of the boundaries of $p_a$ and $p_b$.
         \Statex \quad Concatenate the remaining boundary segments.
         \Statex \quad Traverse this loop to get the ordered cycle $B=(\mathbf{r}_1,\ldots,\mathbf{r}_m)$.
           
    \State \textit{Convexity test (indices modulo $m$):} define $\Delta\mathbf{r}_i:=\mathbf{r}_{i+1}-\mathbf{r}_i$, then compute
    \[
      c_i=(\Delta r_i)_x\,(\Delta r_{i+1})_y-(\Delta r_i)_y\,(\Delta r_{i+1})_x .
    \]
    
    \State \textit{Acceptance rule:} if either $c_i>0$ for all $i$ or $c_i<0$ for all $i$,
           \emph{accept} the merge; otherwise, \emph{skip} this edge.
    \State If accepted, create merged polygon $P$ with vertex cycle $B$; replace $p_a,p_b$ by $P$;
           set $\texttt{merged}\gets 1$.
  \EndFor
  \State Form $\mathcal{M}^{j-1}$ from the current cell set (no hanging nodes; shared-edge conformity).
  \State If $\texttt{merged}=0$, terminate the loop over $j$ (no further coarsening possible).
\EndFor

\end{algorithmic}
\end{algorithm}

Algorithm~1 outlines the default (non-adaptive) coarsening procedure. The proposed coarsening mechanism also admits an adaptive variant where at the first or second coarsening levels the default (non‑adaptive) scheme is still used to perform edge‑based pairwise merges (an element is merged with exactly one adjacent neighbor whenever the union across their shared edge is convex) but at deeper levels (within regions targeted for further reduction) the adaptive scheme permits multi‑neighbor merges in a single step (provided the merged polygon remains convex). This accommodates cases where no pairwise merge is possible, for example when an octagonal region is partitioned into three polygons that violate convexity under any pairwise merge but form a valid larger polygon when merged together. Testing such combinations can be significantly  cheaper at coarser levels because of the reduced number of elements. In practice, we employ both the standard and the adaptive variants, guided by a priori geometric information. As illustrated in Section IV.B, applying the adaptive strategy around certain geometrical features reduces the total computational load.
Fig.~3 and Fig.~4 present uniform and adaptive finest-level examples, respectively, along with their coarsening-based hierarchies. These examples are used in Section~\ref{sec:res}. 

\begin{figure}[t]
  \centering
  \includegraphics[width=0.6\linewidth]{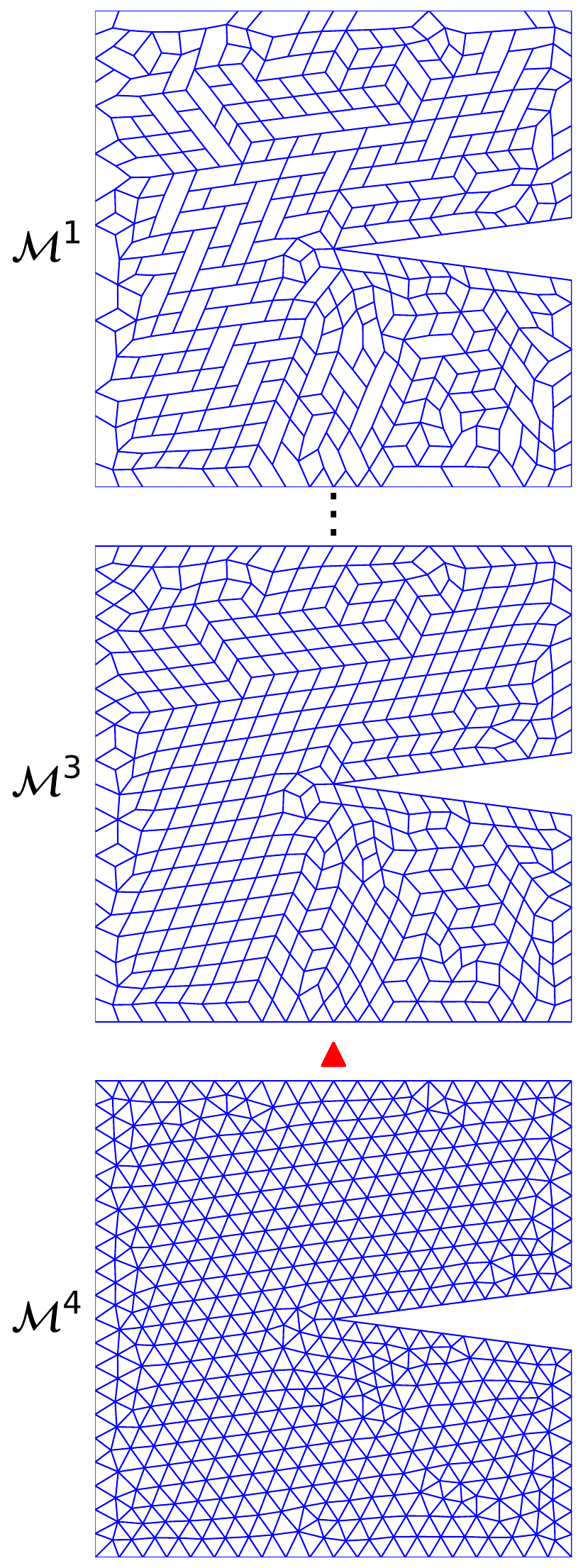} 
    \caption{Illustration of a mesh hierarchy produced by the proposed coarsening algorithm for the wedge scattering problem. From bottom to top: $\mathcal{M}^4$ (input finest-level uniform triangular mesh), $\mathcal{M}^3$ (obtained by coarsening the finest mesh once), and  $\mathcal{M}^1$ (coarsest).}
    \label{fig:mesh-hierarchy-meta}
\end{figure}

\begin{figure}[t]
  \centering
  \includegraphics[width=0.6\linewidth]{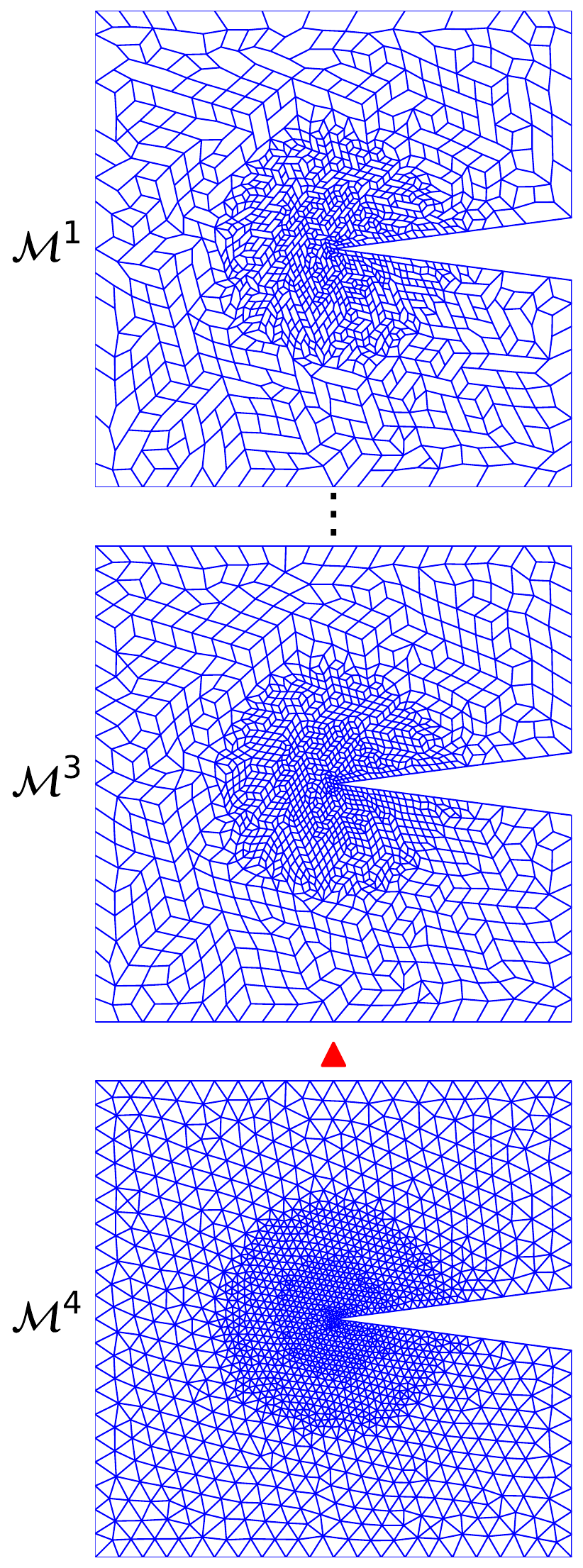} 
  \caption{Illustration of a mesh hierarchy produced by the proposed coarsening algorithm for the wedge scattering problem. From bottom to top: $\mathcal{M}^4$ (input finest-level adaptive triangular mesh), $\mathcal{M}^3$ (obtained by coarsening the finest mesh once), and  $\mathcal{M}^1$ (coarsest).}
    \label{fig:mesh-hierarchy-meta2}
\end{figure}

\subsection{Efficient Computation of Operator-Agnostic Matrices}

The operator-agnostic refinement matrices $\{\tilde{\mathbf{C}}^{j}\}_{j=1}^{q-1}$, which encode coarsening coefficients linking operator-agnostic basis functions between levels $j$ and $j{+}1$ on convex polygonal elements, and their null spaces $\{\tilde{\mathbf{W}}^{j}\}_{j=1}^{q-1}$ (refinement-kernel matrices). These matrices are constructed using Whitney one-form edge basis functions defined on the meshes generated by the mesh coarsening procedure detailed in Subsection~A. They serve as the inputs for the proposed sparse operator-adapted wavelet decomposition algorithm.

To construct the operator-agnostic $\{\tilde{\mathbf{C}}^{j}\}_{j=1}^{q-1}$ matrices efficiently, we first assemble the subdivision matrices $\{\tilde{\mathbf{R}}^{j}\}_{j=1}^{q-1}$ that connects levels $j$ and $j{+}1$. Leveraging the compact support of edge basis functions, $\{\tilde{\mathbf{R}}^{j}\}_{j=1}^{q-1}$ is formed from local inner product vectors and local Gram matrices. For each coarse edge $c$, we define the local inner product vector $\tilde{\mathbf{I}}_{\mathrm{loc}}$ and the local Gram matrix $\tilde{\mathbf{G}}_{\mathrm{loc}}$ associated with the coarse–fine pairs $(\tilde{\boldsymbol{\phi}}_{c}^{j},\,\tilde{\boldsymbol{\phi}}_{f}^{j+1})$ as
\begin{equation}
\renewcommand{\arraystretch}{2} 
\setlength{\arraycolsep}{3pt} % Adjust column spacing
{\tilde{\mathbf{I}}_{\textit{local}}} = \begin{bmatrix}
\langle \tilde{\boldsymbol{\phi}}_{c}^j, \tilde{\boldsymbol{\phi}}_{1}^{j+1} \rangle \\
\langle \tilde{\boldsymbol{\phi}}_{c}^j, \tilde{\boldsymbol{\phi}}_{2}^{j+1} \rangle \\
\vdots \\
\langle\tilde{\boldsymbol{\phi}}_{c}^j, \tilde{\boldsymbol{\phi}}_{\gamma}^{j+1} \rangle
\end{bmatrix}
\end{equation}

and

\begin{equation}
{\tilde{\mathbf{G}}_{\textit{local}}} =
\left[
\begingroup
\renewcommand{\arraystretch}{2} 
\setlength{\arraycolsep}{3pt} % Adjust column spacing
\begin{array}{cccc}
\langle \tilde{\boldsymbol{\phi}}_1^{j+1}, \tilde{\boldsymbol{\phi}}_1^{j+1} \rangle & \langle \tilde{\boldsymbol{\phi}}_1^{j+1}, \tilde{\boldsymbol{\phi}}_2^{j+1} \rangle & \cdots & \langle \tilde{\boldsymbol{\phi}}_1^{j+1}, \tilde{\boldsymbol{\phi}}_{{\gamma}}^{j+1} \rangle \\
\langle \tilde{\boldsymbol{\phi}}_2^{j+1}, \tilde{\boldsymbol{\phi}}_1^{j+1} \rangle & \langle \tilde{\boldsymbol{\phi}}_2^{j+1}, \tilde{\boldsymbol{\phi}}_2^{j+1} \rangle & \cdots & \langle \tilde{\boldsymbol{\phi}}_2^{j+1}, \tilde{\boldsymbol{\phi}}_{\gamma}^{j+1} \rangle \\
\vdots & \vdots & \ddots & \vdots \\
\langle \tilde{\boldsymbol{\phi}}_{\gamma}^{j+1}, \tilde{\boldsymbol{\phi}}_1^{j+1} \rangle & \langle \tilde{\boldsymbol{\phi}}_{\gamma}^{j+1}, \tilde{\boldsymbol{\phi}}_2^{j+1} \rangle & \cdots & \langle \tilde{\boldsymbol{\phi}}_{\gamma}^{j+1}, \tilde{\boldsymbol{\phi}}_{\gamma}^{j+1} \rangle
\end{array}
\endgroup
\right]
\end{equation}
$ \text{for } j = 1, 2, 3, \ldots, q-1$. 

The $\gamma \times 1$ inner-product vectors are computed using (11) and this process is repeated for each coarse-level element edge, where $\gamma$ is the total number of fine-level edges that interact with the selected coarse edge within its associated convex polygons (local computation). The local subdivision vectors $\tilde{\mathbf{R}}_{\mathrm{local}}$ for each coarse edge $c$ are obtained by solving the linear system $\tilde{\mathbf{I}}_{\mathrm{local}} = \tilde{\mathbf{G}}_{\mathrm{local}}\,\tilde{\mathbf{R}}_{\mathrm{local}}$. 
Once all local matrices and vectors have been computed, the global matrix $\tilde{\mathbf{R}}^{j}$ is constructed by local-to-global mapping. Because each selected coarse-level edge interacts only with the fine level edges in its neighboring convex polygons, $\tilde{\mathbf{R}}^{j}$ is very sparse, with only a few nonzero entries in each row and column. Although polygons with more edges at coarser levels increase the bandwidth of the matrices, sparsity is preserved. At the same tine, since smooth regions are covered by fewer polygonal elements, the resulting global matrices are smaller. As seen in the numerical experiments to be described later, these operator-agnostic matrices retain sparsity ratios of $98\%$ or higher on large systems ranging from tens of thousands to millions of unknowns. This sparsity keeps the overall computational complexity near linear. The operator-agnostic matrix $\tilde{\mathbf{C}}^{j}$ is defined as the transpose of the global subdivision matrix.

After computing $\{\tilde{\mathbf{C}}^{j}\}_{j=1}^{q-1}$, we obtain the operator-agnostic matrices $\{\tilde{\mathbf{W}}^{j}\}_{j=1}^{q-1}$ by enforcing $\tilde{\mathbf{C}}^{j}\,\tilde{\mathbf{W}}^{j,T} = \mathbf{0}_{n_j \times N_j}$. Using the highly sparse and banded structure of $\{\tilde{\mathbf{C}}^{j}\}_{j=1}^{q-1}$, we computed $\{\tilde{\mathbf{W}}^{j}\}_{j=1}^{q-1}$ by using the Givens rotation–based QR factorization. This approach computes the null spaces $\{\tilde{\mathbf{W}}^{j}\}_{j=1}^{q-1}$ at near-linear cost, $\mathcal{O}(N)$, and is well suited to parallel implementations. More details on the efficient computation of $\{\tilde{\mathbf{W}}^{j}\}_{j=1}^{q-1}$ are available in~\cite{sik2025oaw,sik2025multiscale}.

\subsection{Numerical Integration and Whitney Forms on Polygonal Elements}

The numerical integrations required to compute the inner products in (11) and (12) are evaluated using the Gaussian quadrature rules. As noted in the literature, numerical integration on arbitrary convex polygons remains an active research topic~\cite{sukumar2006polygonal,sukumar2004numerical, mousavi2010ggq,shivaram2016polygonquad}. In practice, two robust strategies are typically used: (i) triangulation-based quadrature, where the physical polygon is partitioned into triangles and standard Gaussian quadrature rule is applied to each triangle~\cite{sukumar2006polygonal,sukumar2004numerical, mousavi2010ggq,shivaram2016polygonquad,euler2006polygonal,perumal2018brief,cowper1973triangle}; and (ii) direct polygonal cubature, for example, using Green’s theorem or optimized moment-fitting rules that place symmetric integration points inside the polygon. Approaches such as generalized Gaussian quadrature rules for arbitrary polygons, derived via group theory and numerical optimization, are also under active development~\cite{sukumar2006polygonal,wandzura2003triangle,sukumar2004numerical,mousavi2010ggq,ma1996ggq,shivaram2016polygonquad,chen2015minimal}. In this work, we have employed triangulation-based Gaussian quadrature method at all scale levels. This choice is particularly advantageous for our multilevel scheme because the coarse elements are generated by coarsening the finest-level triangular mesh. As a result, at every level of the hierarchy, each coarse polygonal element inherits a consistent sub-triangulation at zero extra cost, and no additional partitioning of polygons is needed. Thus, all necessary integrals (such as those needed to compute $\tilde{\mathbf{R}}$) can be evaluated as sums of contributions from the quadrature points of finest-level triangles, thereby preserving the sparsity pattern and implementation efficiency across the entire hierarchy.

In the implementation, we employ $3$-, $5$-, $7$, $11$, or $16$-point Gaussian quadrature rules per finest-level triangle, selected adaptively: deeper hierarchies or higher polynomial content use $16$, $11$, or $7$ points, whereas shallower hierarchies typically use the cheaper $3$ or $5$ point rules with negligible loss of accuracy. In most of our numerical experiments, the \(5\) or \(7\)-point Gaussian quadrature rule provided accurate and robust results efficiently.

To evaluate the required integrals and assemble the operator-agnostic matrices, we define Whitney one-form edge basis functions on convex polygonal elements using generalized barycentric coordinates. Thanks to the coarsening procedure described in Subsection~A, the vertex coordinates and edge connectivity, together with their full scale history, are available at every resolution level without additional processing. Let $P\subset\mathbb{R}^2$ be a convex $n$-gon with vertices $\{v_i\}_{i=1}^n$ listed and a family $\{\lambda_i(x)\}_{i=1}^n$ is a set of \emph{generalized barycentric coordinates} on $P$ (for all $x\in P$) if
\begin{equation}\label{eq:gbar-axioms}
\lambda_i(x)\ge 0,\qquad
\sum_{i=1}^n \lambda_i(x)=1,\qquad
\sum_{i=1}^n \lambda_i(x)\,v_i = x .
\end{equation}
Given such $\lambda_\ell(x)$ and $\lambda_j(x)$, the Whitney one-form edge basis function $\mathbf{W}_{\ell,j}(x)$ attached to the edge $e_{\ell,j}=(v_\ell,v_j)$ is
\begin{equation}\label{eq:polyWhitney}
\mathbf{W}_{\ell,j}(x)
= \lambda_{\ell}(x)\,\nabla \lambda_{j}(x)
- \lambda_{j}(x)\,\nabla \lambda_{\ell}(x).
\end{equation}
To assign a Whitney one-form edge basis function to any edge of a convex polygon, generalized barycentric coordinates on the cell must first be computed. For convex polygons with more than three vertices, there are infinitely many choices that satisfy \eqref{eq:gbar-axioms}, including Wachspress coordinates and mean-value coordinates (MVC). Both choices produce valid Whitney edge bases via~\eqref{eq:polyWhitney} and, in our tests, yield very similar results; however, in our multiresolution setting, the coarsening procedure can occasionally generate polygons with nearly collinear edges or mild geometric distortion at deeper levels, especially in hierarchies with many levels where the coarsest cells may have higher edge counts. Under such conditions, MVC is markedly more robust, whereas Wachspress coordinates may suffer a loss of accuracy. For this reason, we adopt MVC as the default choice in the numerical experiments presented in this paper. On the other hand, Wachspress coordinates can be faster to evaluate in some cases, but this may come at the expense of accuracy. Nevertheless, our numerical experiments show that, for the most part, both choices yield similar results. Further mathematical details on the numerical integration scheme and the polygonal basis functions discussed in this subsection can be found in~\cite{sukumar2006polygonal,sukumar2004numerical, mousavi2010ggq,shivaram2016polygonquad,chen2015minimal,cowper1973triangle,euler2006polygonal,mukherjee2016polygonal,mukherjee2015hierarchical,perumal2018brief,whitney1957git,gillette2016construction,wachspress1975rational,floater2003mvc,floater2006barycentric,floater2014wachspress,ma1996ggq,wandzura2003triangle}.

\subsection{Construction of the Main Algorithm}
%%%\subsection{Construction of the Multiscale Matrix-Free Solver}

This subsection presents a brief overview of the main matrix-free algorithm to generate multiscale operator-adapted wavelet decomposition-based FEM solutions. The full mathematical and algorithmic details were presented in\cite{sik2025oaw,sik2025multiscale}; here we summarize only the essentials aspects for completeness.

The hierarchy is produced by a nested coarsening procedure. Importantly, no intermediate dense matrices are explicitly assembled. The core operator-adapted linear operator family \(\{\mathbb{C}^{j}\}_{j=1}^{q-1}\) is defined by

\begin{equation}
\begin{aligned}
{\mathbb{C}}^{j-1} =
&\left ({\tilde{\mathbf{C}}}^{j-1} {\tilde{\mathbf{C}}}^{j-1,T}\right)^{-1} 
{\tilde{\mathbf{C}}}^{j-1} \Bigg[
{\mathbb{I}} - {\mathbb{A}}^{j} {\tilde{\mathbf{W}}}^{j-1,T} \cdots \\
&  \cdots \left(
{\tilde{\mathbf{W}}}^{j-1} {{\mathbb{A}}^{j} }{\tilde{\mathbf{W}}}^{j-1,T}
\right)^{-1} 
{\tilde{\mathbf{W}}}^{j-1}
\Bigg].
\label{eq:hierarchical_C}
\end{aligned}
\end{equation}
where $\{\mathbb{A}^j\}_{j=1}^{q}$ are defined recursively by $\mathbb{A}^{j}=\mathbb{C}^{j}\mathbb{A}^{j+1}\mathbb{C}^{j,T}$.
Since \(\tilde{\mathbf{A}}^{\,q}=\mathbb{A}^{\,q}\), \(\{\mathbb{C}^{j}\}_{j=1}^{q-1}\) and \(\{\mathbb{A}^{j}\}_{j=1}^{q-1}\) are hierarchically evaluated from the finest to the coarser levels using only sparse operator-agnostic matrices and vectors. The right-hand side at the coarsest level, \(\mathbbm{g}^{\,1}\), and the right-hand side at the detail level, \(\mathbbm{b}^{\,j}\) for \(j=1,\dots,q-1\), are
\begin{equation}
\label{eq:RHS}
\begin{aligned}
\mathbbm{g}^{\,1} &= \mathbb{C}^{1}\mathbb{C}^{2}\cdots \mathbb{C}^{\,q-1}\,\mathbbm{g}^{\,q},\\
\mathbbm{b}^{\,j} &= \tilde{\mathbf{W}}^{\,j}\,\mathbb{C}^{\,j+1}\cdots \mathbb{C}^{\,q-1}\,\mathbbm{g}^{\,q}.
\end{aligned}
\end{equation}
The coefficients \(\mathbbm{v}^{\,1}\) and \(\mathbbm{w}^{\,j}\) are computed from \({\mathbb{A}}^1 {\mathbbm{v}}^1 = {\mathbbm{g}}^1\) and \({\tilde{\mathbf{W}}}^{j} {\mathbb{A}}^{j+1} {\tilde{\mathbf{W}}}^{j,T} {\mathbbm{w}}^j = {\mathbbm{b}}^j\).  Consequently, both coarser and detail-level solutions can be expressed using only sparse matrix-vector multiplications as follows:

\begin{equation}
     {\mathbbm{s}}^{\text{coarse}}   = \tilde{{\mathbf{\Phi}}}^{q,T} {\mathbb{C}}^{q-1,T} {\mathbb{C}}^{q-2,T} \cdots {\mathbb{C}}^{1,T}  {\mathbbm{v}}^1
\end{equation}
\begin{equation}
    {\mathbbm{s}}^{\text{detail, j}}   = \tilde{{\mathbf{\Phi}}}^{q,T} {\mathbb{C}}^{q-1,T} \cdots {\mathbb{C}}^{j+1,T} {\tilde{\mathbf{W}}}^{j, T}  {\mathbbm{w}}^j
\end{equation} 

Throughout, \(\{\mathbb{C}^{j}\}_{j=1}^{q-1}\) and \(\{\mathbb{A}^{j}\}_{j=1}^{q}\) are treated as linear operators rather than in an explicitly assembled matrix form. Although inverse symbols appear in \eqref{eq:hierarchical_C}, they are implemented as inner linear solves and  no explicit matrix inversions are performed.

All linear systems obtained are solved using GMRES Krylov subspace iterative solvers with ILU preconditioning to accelerate convergence\cite{saad2003iterative}. Since the intermediate matrices would be dense if formed explicitly, the ILU factorizations are constructed from sparse mimics of these matrices; a complete description and analysis are provided in\cite{sik2025oaw}. In numerical experiments, unrestarted GMRES with ILU preconditioner typically converged in \(20\)–\(250\) iterations for problems ranging from thousands to millions of unknowns with a residual tolerance of \(\varepsilon=10^{-4}-10^{-5}\).

\section{Numerical Experiments\label{sec:res}}

In this section, the operator-adapted wavelet decomposition–based FEM, constructed on a coarsening-based polygonal mesh hierarchy, is applied to two representative multiscale electromagnetic problems: scattering by a perfectly conducting wedge and a microporous material slab.

\subsection{Scattering by a Perfectly Conducting Wedge}

Scattering of EM waves by a 2D perfectly conducting wedge has received significant attention for many decades \cite{sevgi2014wedge,balanis2012aee}. The wedge serves as a canonical model with tip singularity. This problem is particularly well suited to demonstrate the advantages of the proposed multiscale algorithm because accurately resolving the near-field behavior in the vicinity of the wedge tip using usual Method of Moments (MoM) or FEM formulations typically requires extremely fine meshes, incurring substantial computational overhead.

Using the proposed algorithm, both far-field and near-field behavior can be captured within a single simulation by exploiting scale separation. Coarse levels accurately represent the far-field, whereas fine levels are activated only when needed to resolve the field behavior in the vicinity of the tip/edge of the wedge. Owing to scale decoupling, fine-level solutions can be superposed onto the already computed coarse solution without re-computation. Consequently, the desired accuracy near the tip of the wedge can be achieved by incrementally augmenting the far-field solution with fine-scale detail solutions, all within the same polygonal mesh hierarchy and algebraic framework.

The computed results using the proposed algorithm are validated against the (finest-level) conventional FEM solution, the MoM solution, and the series solution given by~\cite{balanis2012aee}:
\begin{equation}
\label{eq:wedge_series}
E_z^{t} = E_0\sum_{\nu=1}^{\infty} j^{\nu}\, J_{\nu}(\beta\rho)\,
\sin\!\big[\nu(\phi' - \alpha)\big]\,
\sin\!\big[\nu(\phi - \alpha)\big],
\end{equation}
for the total field produced by an incident TM$^z$ uniform plane wave with amplitude $E_0$. In Eq.~(\ref{eq:wedge_series}), $J_{\nu}(\cdot)$ denotes the Bessel function of the first kind of order $\nu$; $\rho$ is the radial distance from the wedge tip; $\beta$ is the wavenumber; $\nu = n\pi\big/ \big(2\pi\alpha\big)$ with $n=1,2,\ldots$; and \(\phi\) and \(\phi'\) are the observation and incidence (illumination) angles, respectively. Angles are measured counterclockwise with $0^\circ$ along the wedge bisector; the wedge faces lie at $\pm\alpha$ degrees.

\begin{figure}[t]
  \centering
\includegraphics[width=1\linewidth]{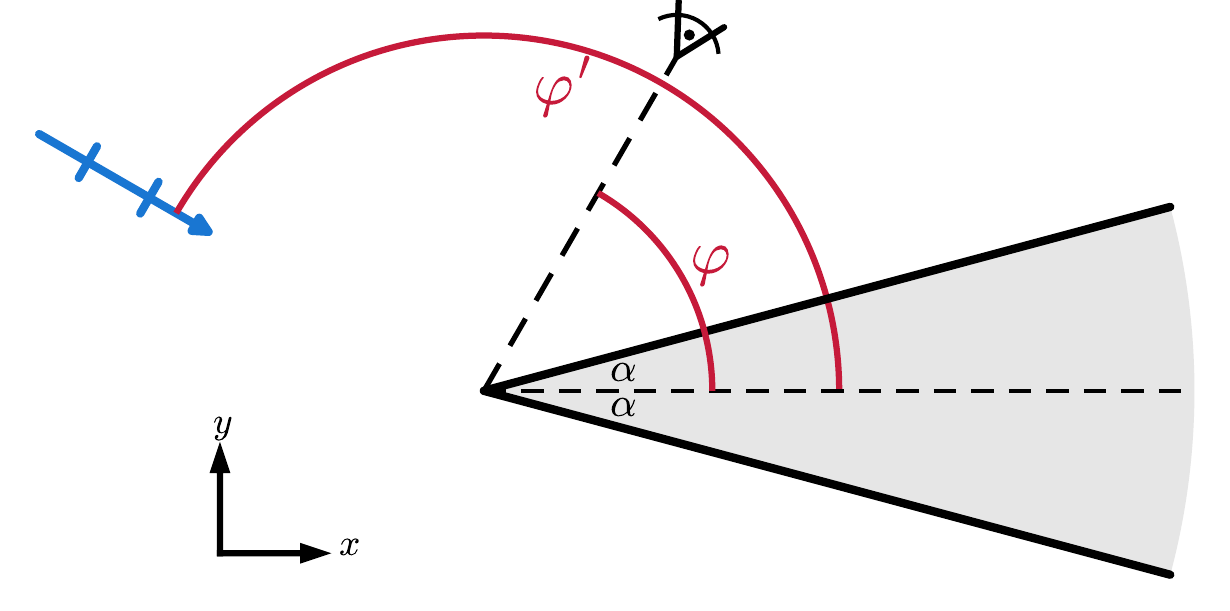} 
  \caption{Geometry for the PEC wedge–scattering example. A perfectly conducting wedge with interior (opening) angle \(2\alpha\) is illuminated by a plane wave incident at angle \(\phi'\). The observation angle \(\phi\) is measured counterclockwise from the wedge bisector (\(0^\circ\)); the wedge faces lie at \(\phi=\pm\alpha\) degrees. Axes indicate the 2D \(x\)–\(y\) plane.}
  \label{fig:wedge_geo}
\end{figure}

\begin{figure}[t]
  \centering
  \includegraphics[width=0.65\linewidth]{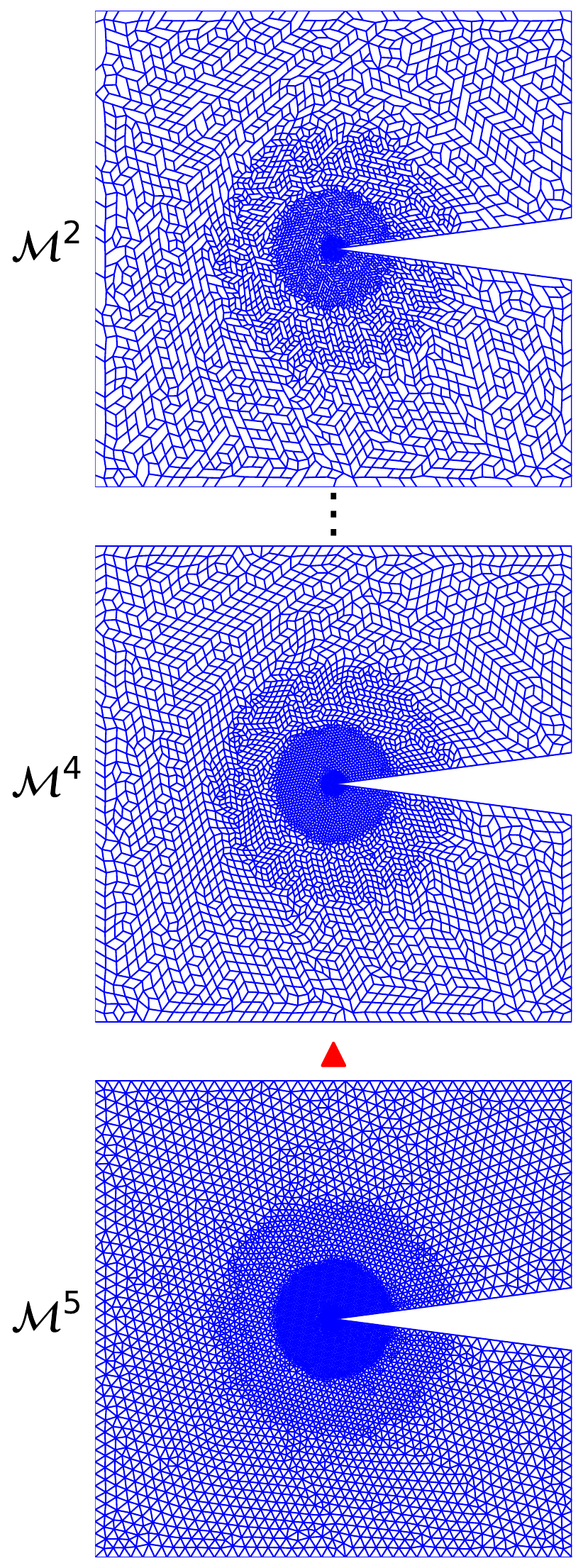}
  \caption{Mesh hierarchy generated by the proposed coarsening algorithm for the wedge–scattering numerical experiment (\(q=5\)). From bottom to top: \(\mathcal{M}^{5}\) (input finest–level adaptive triangular mesh), \(\mathcal{M}^{4}\) (finest mesh coarsened once), and \(\mathcal{M}^{2}\) (finest mesh coarsened three times).}
  \label{fig:mesh-hierarchy-wedgefine}
\end{figure}

First, far-field patterns obtained from the exact series solution, MoM, conventional FEM, and the operator-adapted wavelet–decomposition FEM are presented. In these numerical experiments, the finest-level mesh is kept coarser than in the near-field accuracy studies discussed later. Fig.~7 shows these far-field radiation patterns for different incidence and wedge opening angles where excellent agreement is observed even on comparatively coarse meshes. 

Accurately resolving the near field in the immediate vicinity of the wedge tip with MoM or conventional FEM requires exceedingly fine discretization. The subsequent discussion focuses on the near-field investigation for the geometry shown in Fig.~5.

\begin{figure}[t]
  \centering
  \begin{subfigure}{0.5\linewidth}
    \centering
    \includegraphics[width=\linewidth]{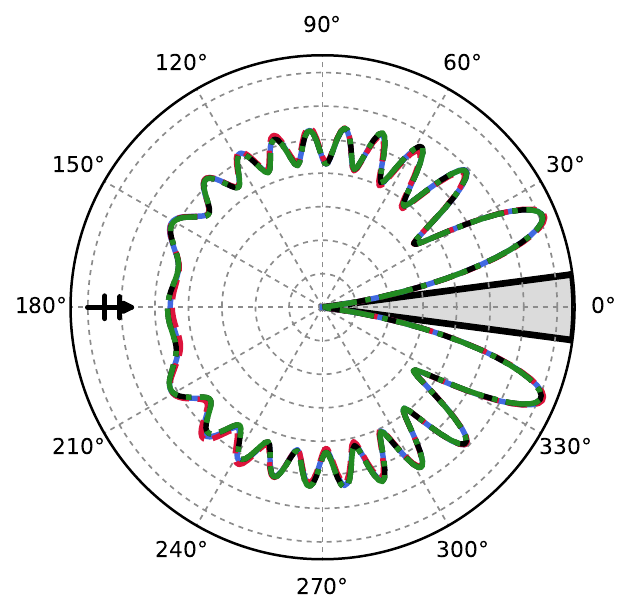}
    \caption{}
  \end{subfigure}\hfill
  \begin{subfigure}{0.5\linewidth}
    \centering
    \includegraphics[width=\linewidth]{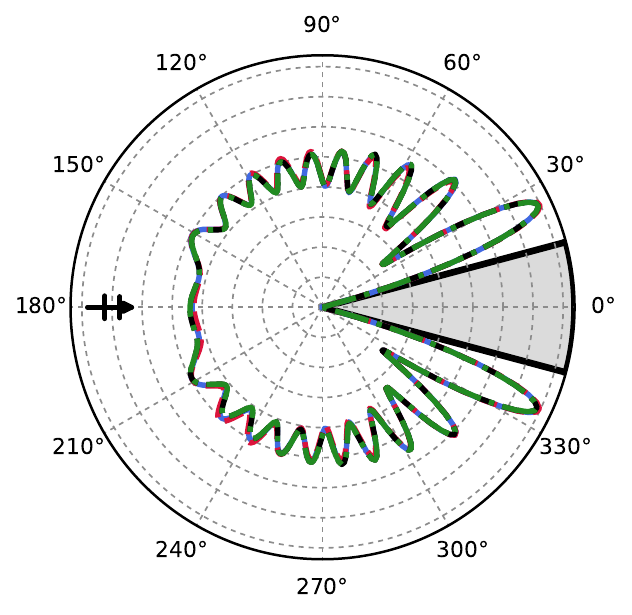}
    \caption{}
  \end{subfigure}

  \vspace{0.35em}

  \begin{subfigure}{0.5\linewidth}
    \centering
    \includegraphics[width=\linewidth]{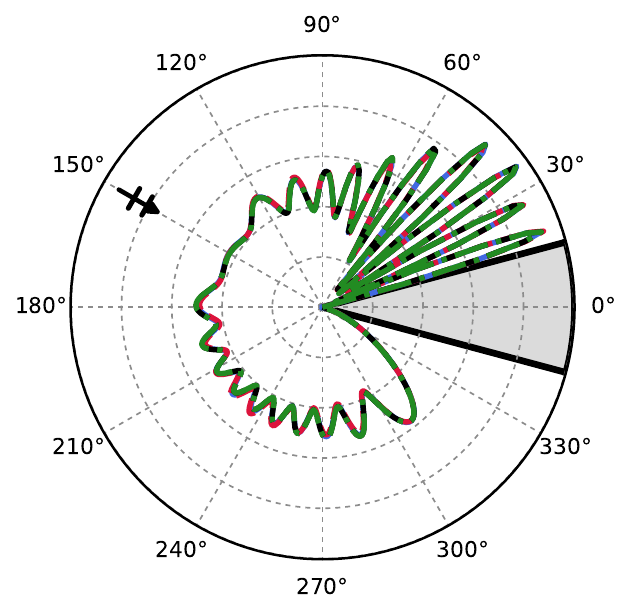}
    \caption{}
  \end{subfigure}\hfill
  \begin{subfigure}{0.5\linewidth}
    \centering
    \includegraphics[width=\linewidth]{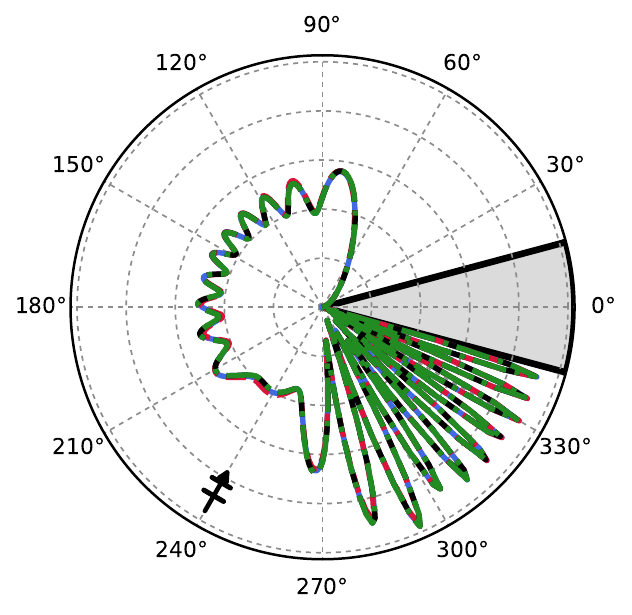}
    \caption{}
  \end{subfigure}
  \vspace{0.5em}
  \includegraphics[width=1\linewidth]{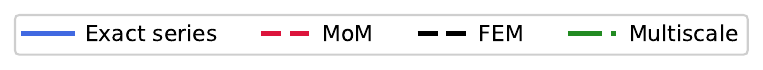}
  \caption{Far-field radiation patterns for scattering by a perfectly conducting wedge, shown for four methods and four scenarios: (a) \( \phi' = 180^\circ,\ \alpha = 7.5^\circ \);
  (b) \( \phi' = 180^\circ,\ \alpha = 15^\circ \);
  (c) \( \phi' = 150^\circ,\ \alpha = 15^\circ \);
  (d) \( \phi' = 240^\circ,\ \alpha = 15^\circ \).}
  \label{fig:four-panels-with-shared-legend}
\end{figure}

In the near-tip area, accuracy is evaluated over the region defined by a circle of radius \(\lambda/10\) centered at the wedge tip. To reach a relative \(L^2\)-norm error of about \(1.5\%\) in this region using conventional FEM, element edge lengths on the order of \(\lambda/300\)–\(\lambda/350\) are required; the error is computed as the relative \(L^2\) norm between the conventional FEM solution and the series solution. For a robust reference, the exact series solution is computed with \(N=500\) terms. In the example under discussion, a TM$^z$ plane wave at \(30\,\mathrm{MHz}\) with incidence angle \(\phi'=180^\circ\) illuminates the wedge, and error calculations are restricted to the near-tip region, defined as a circle of radius \(\lambda/10\) centered at the tip. 

The adaptive finest-level mesh and the coarsening-based polygonal mesh hierarchy are shown in Fig.~6, while Fig.~8 illustrates the solutions across different scale levels. At the finest level, the mesh employs triangles with edge lengths of approximately \(\lambda/300\) within the \(\lambda/10\) neighborhood of the tip, with the element size increasing gradually with distance up to about \(\lambda/50\) in the far-field. For the five-level hierarchy, two level solution (the superposition of the coarsest level and the first detail level) nearly reproduces the far-field behavior; however, near-field features around the wedge tip are largely absent. After adding a second detail level, the far-field becomes almost indistinguishable from the analytic reference and near-field tip-adjacent features begin to emerge. With the third detail level, near-tip features are clearly resolved, and the field closely matches the reference series solution across almost the entire domain, except for a very small region directly in front of the tip. Including the fourth detail level recovers the remaining tip-region details; the relative \(L^2\)-norm error, evaluated within the \(\lambda/10\) near-tip circle and measured against the finest-level usual FEM solution, falls below \(2\%\). When benchmarked against the reference series solution, the relative \(L^2\)-norm error in the same region is approximately \(3\%\) or less. This behavior illustrates that coarse levels efficiently deliver the far-field, while near-tip physics are recovered by activating detail levels on the same hierarchy. Unlike conventional FEM, our approach avoids solving very large linear systems.

The adaptive polygonal mesh hierarchy covers smooth regions with fewer, larger, high-sided coarse polygonal elements, whereas high-resolution areas are captured by using finer elements. The resulting linear systems are much smaller—cutting peak memory by as much as\(\ 40\%\).

For configurations targeting under\(\ 3\%\) relative \(L^2\)-norm error relative to the finest-level conventional FEM solution) using three scale levels, the polygonal element compositions at each level are as follows:

\begin{itemize}
  \item $\mathcal{M}^{1}$: 46 triangles, 6106 quadrilaterals, 132 pentagons, 1848 hexagons, 88 octagons, 18 decagons, and 3 dodecagons;  
  \item $\mathcal{M}^{2}$: 279 triangles, 13710 quadrilaterals, 3 pentagons, and 1 hexagon;
  \item $\mathcal{M}^{3}$: 29730 triangles.
\end{itemize} 

\begin{figure}[t]
  \centering
  \begin{subfigure}{0.5\linewidth}
    \centering
    \includegraphics[width=\linewidth]{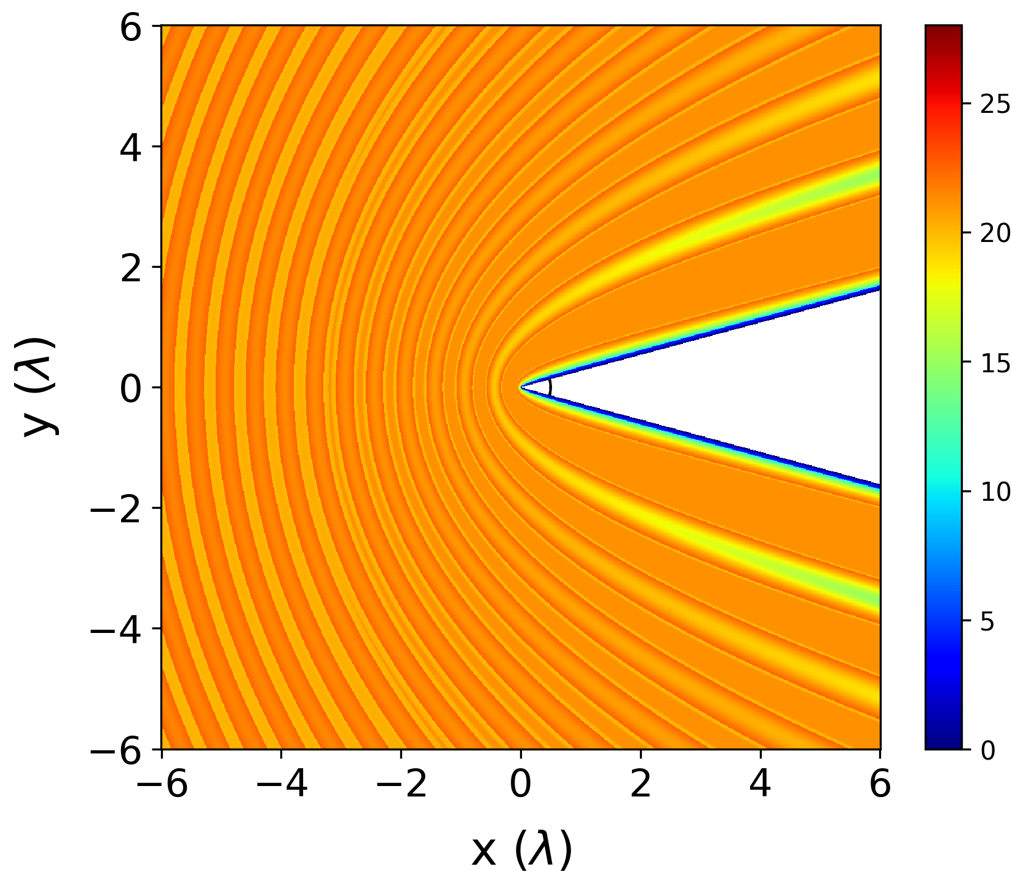}
    \caption{}
  \end{subfigure}\hfill
  \begin{subfigure}{0.5\linewidth}
    \centering
    \includegraphics[width=\linewidth]{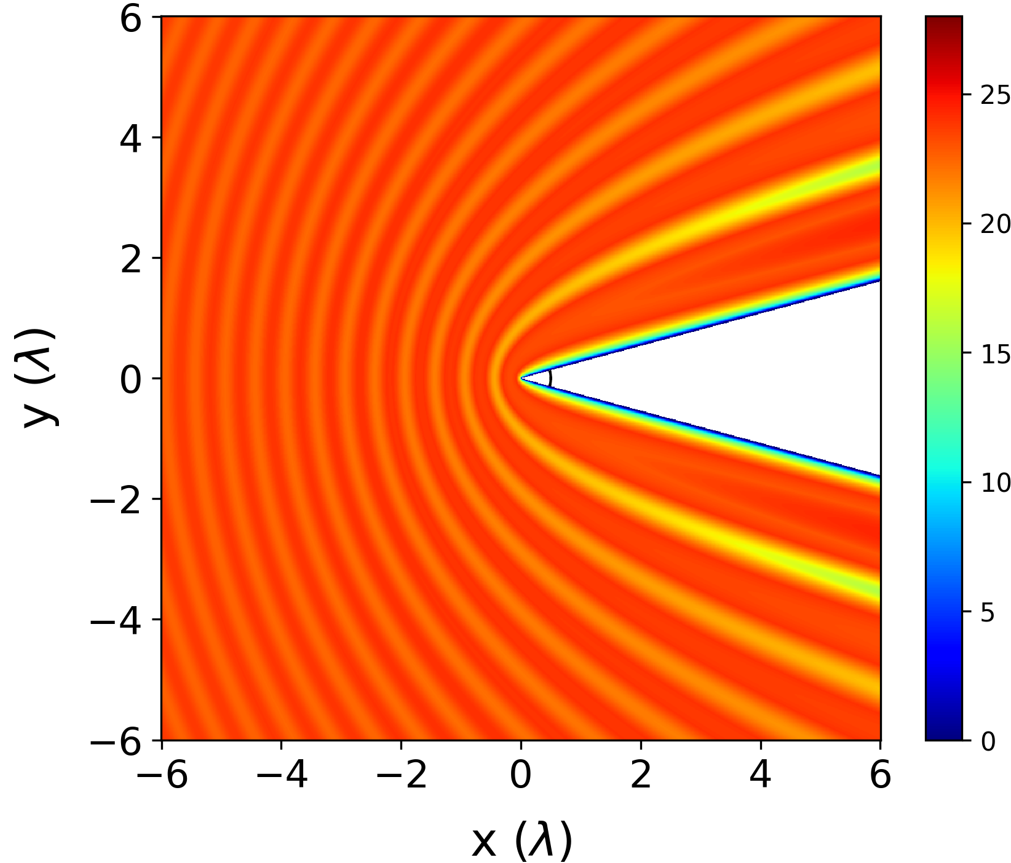}
    \caption{}
  \end{subfigure}

  \vspace{0.35em}

  \begin{subfigure}{0.5\linewidth}
    \centering
    \includegraphics[width=\linewidth]{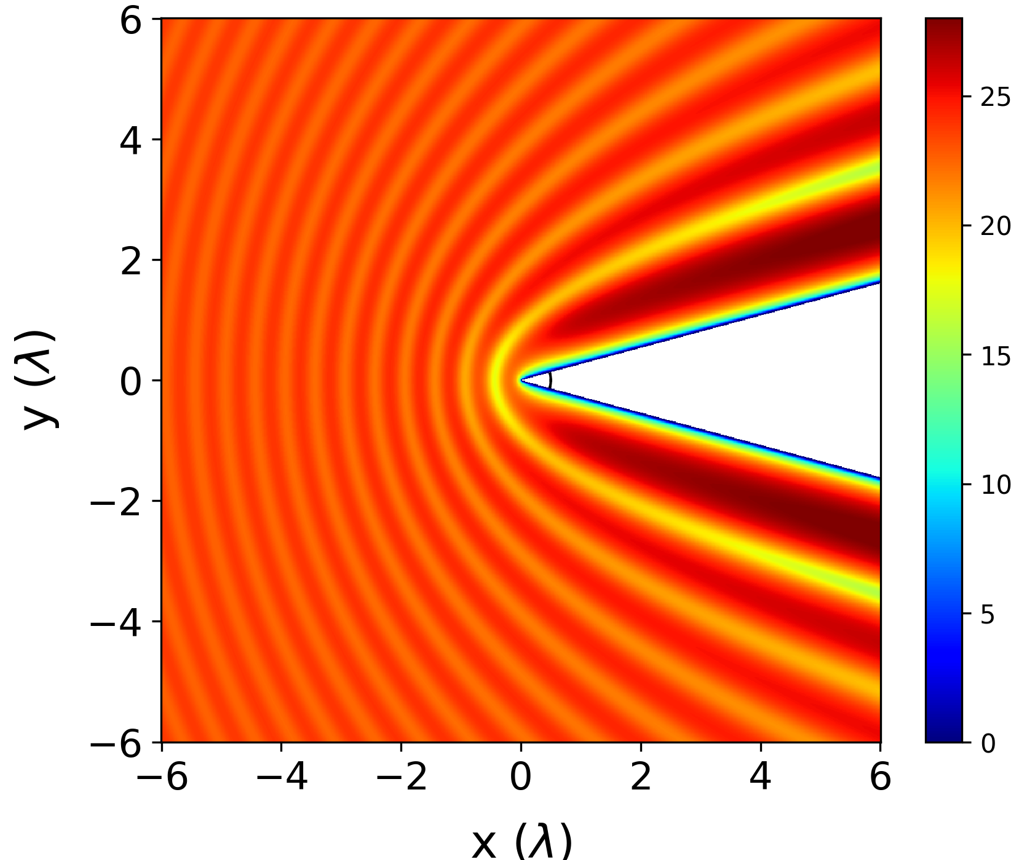}
    \caption{}
  \end{subfigure}\hfill
  \begin{subfigure}{0.5\linewidth}
    \centering
    \includegraphics[width=\linewidth]{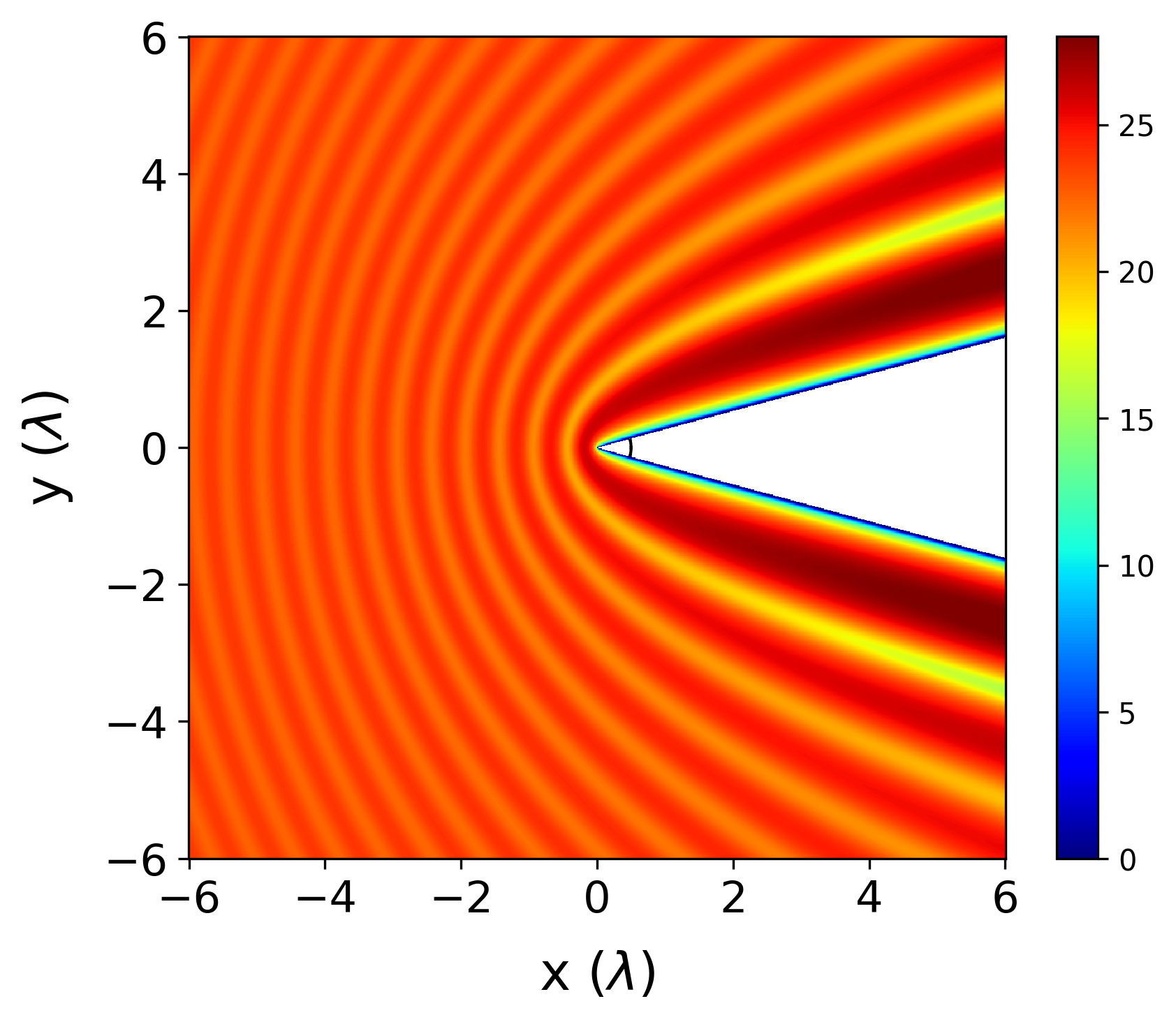}
    \caption{}
  \end{subfigure}
  \caption{Computed electric field magnitude $\mathbbm{E}$ for the scattering by a perfectly conducting wedge using the operator-adapted wavelet decomposition FEM with the proposed coarsening-based polygonal mesh hierarchy: (a) Two-level solution, \({\mathbbm{E}} = {\mathbbm{E}}^{\text{coarse}} + {\mathbbm{E}}^{\text{detail, 1}}\), (b) Three-level solution,  (c) Four-level solution, and (d) Five-level solution, \({\mathbbm{E}} = {\mathbbm{E}}^{\text{coarse}} + {\mathbbm{E}}^{\text{detail, 1}}+{\mathbbm{E}}^{\text{detail, 2}}+{\mathbbm{E}}^{\text{detail, 3}}+{\mathbbm{E}}^{\text{detail, 4}}.\)}
\end{figure}

Fig.~9 presents a log-log plot of CPU time per iteration \(T\) versus degrees of freedoms (DoFs) \(N\); the slope of the \(\log T\)--\(\log N\) fit for the coarsest-level calculation is approximately 0.94, indicating near-linear cost. Once \(\mathbbm{s}^{\text{coarse}}\) is computed, finer-level solutions can be obtained very efficiently: computing \(\mathbbm{s}^{\text{detail,1}}\) costs about \(\mathcal{O}(N^{0.59})\), \(\mathbbm{s}^{\text{detail,2}}\) about \(\mathcal{O}(N^{0.37})\), \(\mathbbm{s}^{\text{detail,3}}\) about \(\mathcal{O}(N^{0.23})\), and \(\mathbbm{s}^{\text{detail,4}}\) about \(\mathcal{O}(N^{0.11})\). Moreover, when precomputation costs are included, the costs remains near-linear: the global \(\log T\)-\(\log N\) fit exhibits a slope of approximately \(1.04\). The complexity trends are validated across a wide range of scales, with the number of DoFs at the finest level ranging from thousands to millions.

\begin{figure}[H]
    \centering
    \includegraphics[width=1\linewidth]{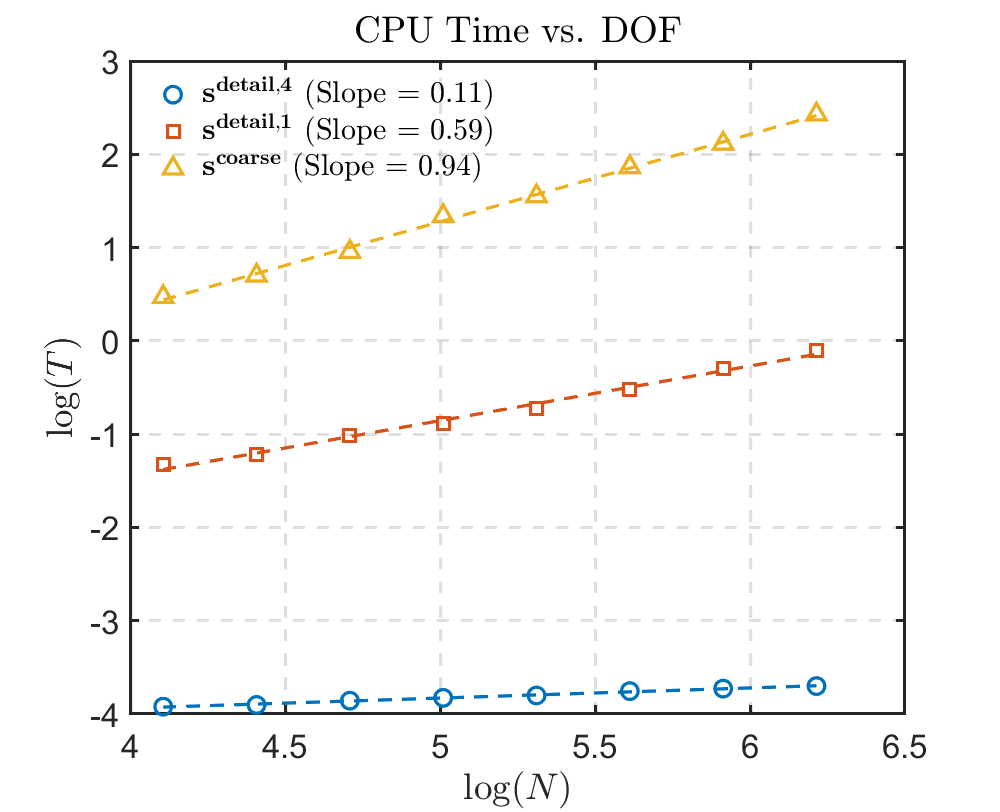}
    \captionsetup{width=1\linewidth} 
     \caption{Elapsed time per iteration versus DoF for the coarsest ($\mathbbm{s^{\text{coarse}}}$), first detail ($\mathbbm{s^{\text{detail,1}}}$), and fourth detail ($\mathbbm{s^{\text{detail,4}}}$)  level solutions for the five-level scenario.}
\end{figure}

\subsection{MPSi slab superlensing effect}

In this subsection, a microporous Si (MPSi) slab problem is investigated using the proposed multiscale algorithm. This type of slab exhibits the \emph{superlensing effect}, i.e., spot–size focusing below the classical diffraction limit. The problem was previously analyzed using FDTD-based approaches in~\cite{donderici2005subgridding, valverde2025TAP, luo2002aanr,li2003lensing}. 
\begin{figure}[H]
  \centering
  \includegraphics[width=1\linewidth]{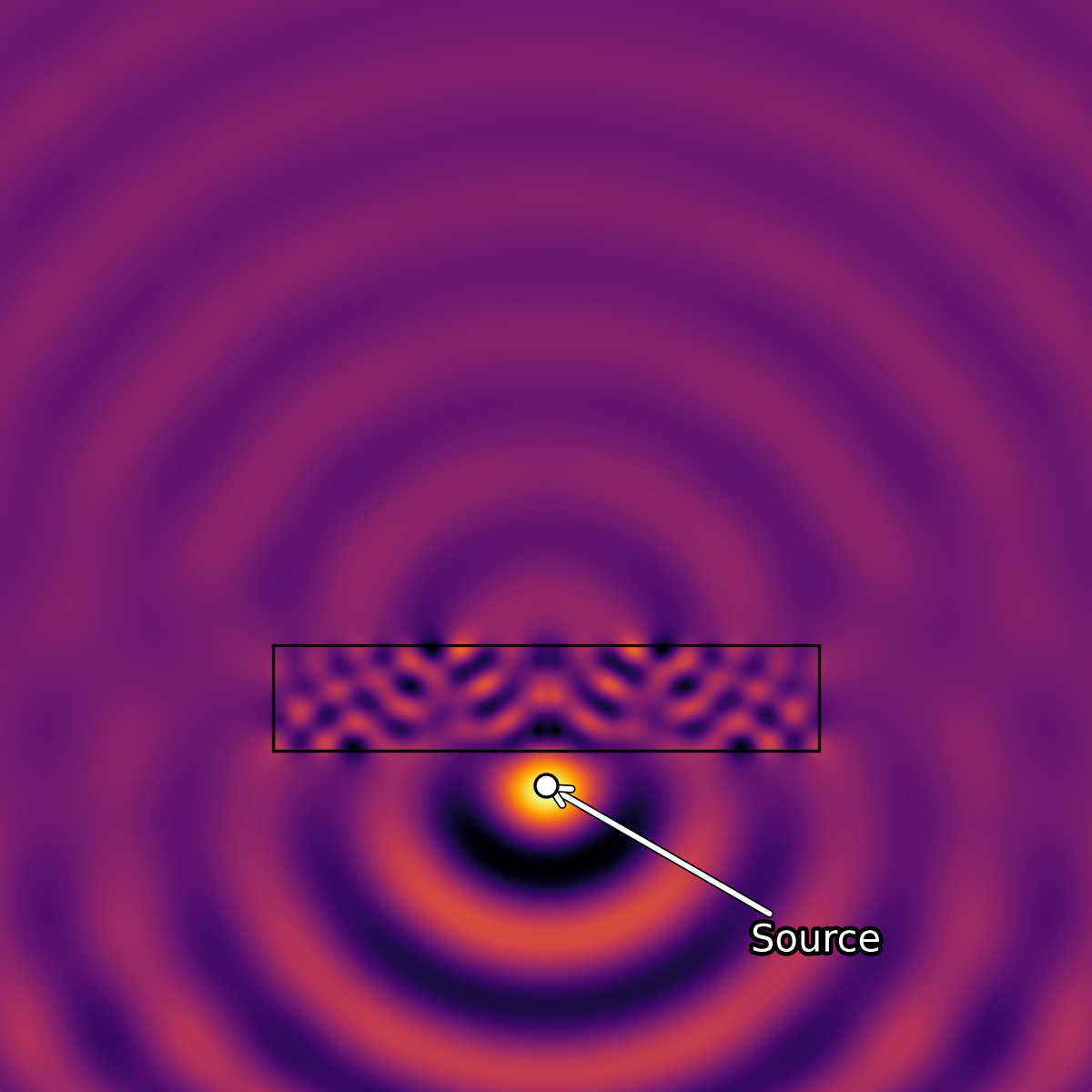}
  \captionsetup{width=1\linewidth}
  \caption{Electric-field distribution for the microporous Si slab example obtained with the five-level multiscale solution, computed as $\mathbbm{E}=\mathbbm{E}^{\text{coarse}}+\sum_{k=1}^{4}\mathbbm{E}^{\text{detail},\,k}$.}
\end{figure}
Fig.~10 shows the electric field distribution computed using the operator–adapted wavelet–decomposition FEM with a five scale level hierarchy, for a point source placed next to the MPSi slab perforated with a rectangular array of cylindrical air holes. In this scenario, the superlensing effect is observed: a distinct image forms on the opposite side of the MPSi slab. In the geometry of Fig.~10, the diameter of the holes is \(2a=0.125\,\lambda_0\) and the lattice period is \(h=0.18333\,\lambda_0\), where \(\lambda_0\) denotes the free–space wavelength.

\begin{figure}[t]
  \centering
  \includegraphics[width=0.9\linewidth]{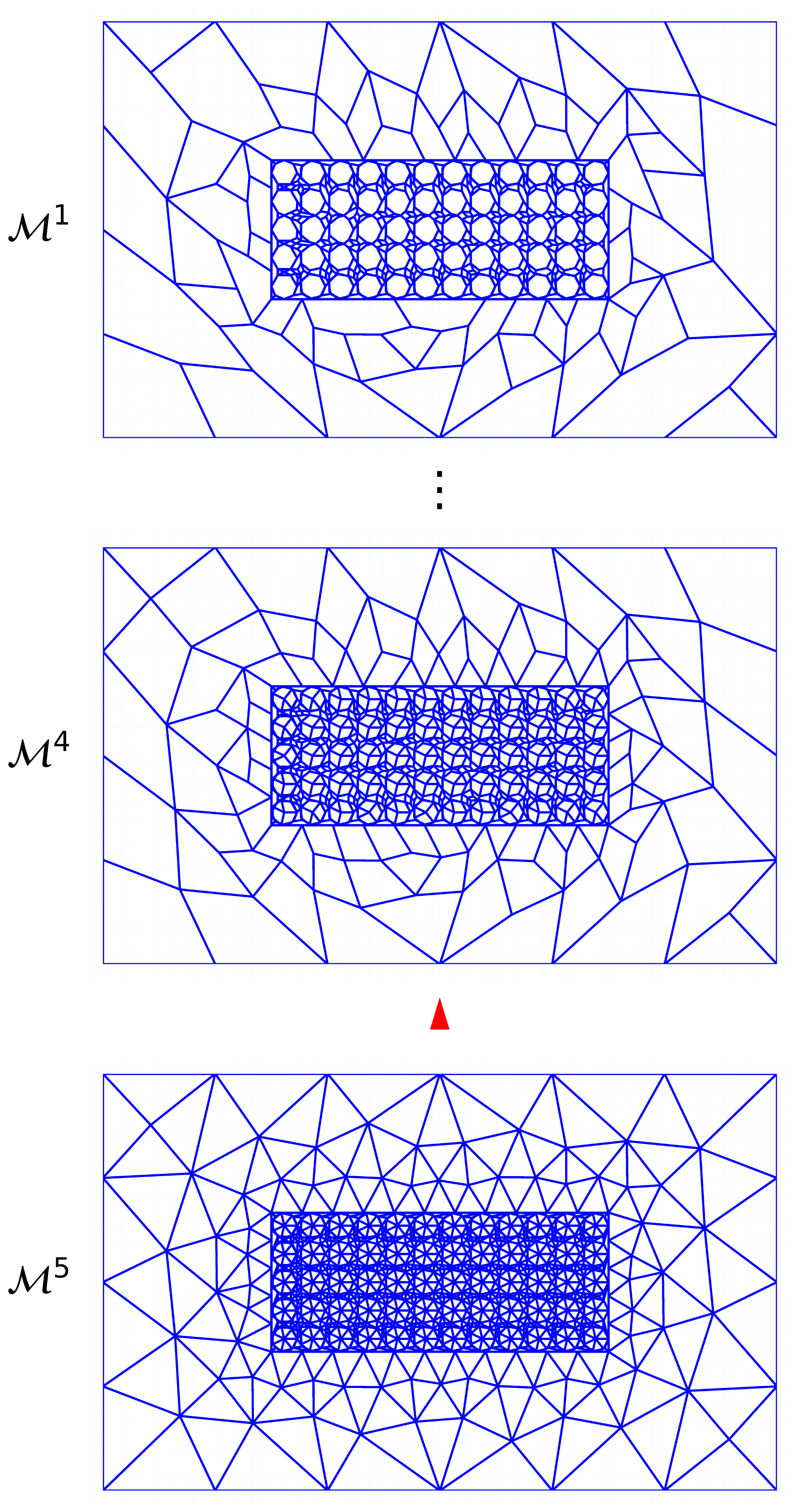} 
  \caption{Mesh hierarchy generated by the proposed coarsening procedure with $q=5$ levels for the MPSi slab example. From bottom to top: $\mathcal{M}^5$ (input finest-level adaptive triangular mesh), $\mathcal{M}^4$ (obtained by coarsening the finest mesh once), and $\mathcal{M}^1$ (coarsest level mesh).}
  \label{fig:mesh-hierarchy-meta}
\end{figure}
Using the proposed multiscale approach, the computed fields show excellent agreement with both usual (adaptive or uniform) finite element implementations and reference FDTD results reported in the literature\cite{donderici2005subgridding, valverde2025TAP, luo2002aanr,li2003lensing}. Beyond accuracy, this algorithm maintains near–linear computational complexity from the precomputation steps through the final linear solves. Moreover, when the coarsening‑based five‑level adaptive mesh hierarchy of Fig.~11 is employed, its polygonal element composition by level is as follows:

\begin{itemize}
  \item $\mathcal{M}^{1}$: 53 triangles, 198 quadrilaterals, 34 pentagons, 23 hexagons, 23 heptagons, 23 octagons, and 11 nonagons;
  \item $\mathcal{M}^{4}$: 151 triangles, 667 quadrilaterals;
  \item $\mathcal{M}^{5}$: 1485 triangles.
\end{itemize}

By employing the proposed polygonal coarsening procedure, peak memory usage falls below 40\% compared with the conventional FEM.
Furthermore, as depicted in Fig.~11, the proposed adaptive coarsening implementation enables incorporating a priori geometric information. For instance, as illustrated in Fig.~11, the air holes at the coarsest level appear as multi-sided polygons (e.g., heptagons, octagons, or nonagons). This approach allows for a further reduction in the number of elements at coarser levels.

\section{Conclusion \label{sec:conclusion}}

This study introduced a sparse operator–adapted wavelet–decomposition finite–element framework with a coarsening–based mesh hierarchy incorporating convex polygonal elements, thereby combining multiscale modeling with localized mesh adaptivity.  In this setting, smooth regions are represented by fewer, larger, polygons, whereas high–gradient neighborhoods retain fine resolution by using smaller elements. This yields a favorable accuracy–cost–memory trade-off: coarse levels capture the response of the smooth regions efficiently, while detail levels are activated only when needed, for example, to resolve evanescent waves, near-field effects, or the physics around geometric discontinuities and/or very fine features.

From an algorithmic standpoint, another key advance with respect to the previous subdivision-based approach developed in~\cite{sik2025oaw,sik2025multiscale} is how the precomputed operator-agnostic matrices for convex polygonal elements are constructed based on Whitney one-forms defined via generalized barycentric coordinates and by using robust quadrature rules over the inherited subtriangles within each polygon. Crucially, scale decoupling, sparsity, and near–linear cost of the method are preserved, which is confirmed by numerical experiments. The perfectly conducting wedge experiment showcases the ability of the proposed method to better capture the singular field behavior near the tip by activating detail levels solutions and solving small independent linear systems for each level. The MPSi slab experiment showcases the ability of the algorithm to exploit polygon elements to render complex geometries with less elements. These capabilities, among others, make the proposed method a compelling tool for multiscale EM analysis.

\appendices


\ifCLASSOPTIONcaptionsoff
  \newpage
\fi



%\bibliographystyle{IEEEtran}


\bibliographystyle{ieeetr} 
\bibliography{main}   


%\vfill



\end{document}